\documentclass[12pt,twoside]{article}
\usepackage{fleqn,espcrc1}

\newcommand {\be}{\begin{equation}}
\newcommand {\eb}{\end{equation}}
\newcommand{\ba}{\begin{eqnarray}}
\newcommand{\ea}{\end{eqnarray}}
\newcommand{\pp}{$\pi\pi$ }
\newcommand{\roro}{$\sigma\sigma$ }
\newcommand{\kk}{$K\overline{K}$ }
\newcommand{\fo}{$f_0(980)$ }
\newcommand{\epw}{$f_0(1400)$ }
\newcommand{\epsig}{$f_0(500)$ }

\newcommand{\reactpol}{$\pi^- p_{\uparrow} \rightarrow \pi^+ \pi^- n$ }

\newcommand{\AmS}{{\protect\the\textfont2
  A\kern-.1667em\lower.5ex\hbox{M}\kern-.125emS}}

\title{Properties of scalar--isoscalar mesons from
multichannel interaction analysis below 1800 MeV}

\author{R. Kami\'nski
$^{\mbox{\scriptsize a, b}}$,
L. Le\'sniak 
\address{Department of Theoretical Physics,
The Henryk Niewodnicza\'nski Institute of Nuclear Physics,
 PL 31-342 Krak\'ow, Poland}
and
B.\ Loiseau\address{LPTPE Universit\'e P.~et M.~Curie,
4, Place Jussieu, 75252 Paris CEDEX 05, France}}
      
\begin{document}

\maketitle

\begin{abstract}
Scalar-isoscalar mesons
 are studied using an unitary model
in three channels: $\pi\pi$, \kk and an
effective $2\pi2\pi$.
All the solutions, fitted to the \pp and \kk data, exhibit 
a wide $f_0(500)$, a narrow \fo and two relatively 
narrow resonances, lying on different sheets
between  1300 MeV and 1500 MeV. These latter states are similar to 
the $f_0(1370)$ and $f_0(1500)$ seen in experiments at CERN.
Branching ratios are compared with
available data. 
We have started investigations of some crossing symmetry and chiral constraints
imposed near the \pp threshold on the scalar-isoscalar, scalar-isotensor and
P-wave \pp amplitudes.
\end{abstract}

\section{INTRODUCTION}
Study of scalar-isoscalar mesons is an important issue of QCD
: one expects presence of some scalar
($J^{PC} = 0^{++}, I=0$) glueballs~\cite{pdg98}
. Here we shall  try to see what
 can be learned from the
present experimental knowledge of the scalar-isoscalar \pp and \kk
phase shifts.  We consider an unitary model  with
separable interactions  in three channels: $\pi\pi$, \kk and an effective
$2\pi2\pi$, denoted \roro, in a mass range from the \pp threshold up to 1800
MeV \cite{kll297,kll299}.
Several solutions are obtained by fitting \pp phase shifts 
from the CERN-Cracow-Munich analysis
of the \reactpol reaction on a polarized target
\cite{klr} together with lower energy \pp  
and \kk data from reactions on unpolarized target 
(see references given in~\cite{kll299}).

\section{RESULTS}

The different solutions A, B, E and F of our model
are characterized by  presence or absence of \kk and \roro bound states
 when all the interchannel couplings are switched off (see
Table 2 of ~\cite{kll299}). For the fully coupled case, poles of the
$S$-matrix, located in the complex energy plane not
too far from the physical region, are interpreted as
scalar resonances. In all our solutions we find a wide $f_0(500)$,
a narrow \fo and a relatively
narrow $f_0(1400)$ which splits into
 two resonances, lying on different sheets classified according to the sign
 of $Imk_{\pi \pi}, Imk_{K \overline{K}}, Imk_{\sigma \sigma}$.
 Their average masses and widths are
 summarized in Table \ref{resonances}. The finding of  the
 two states near 1400 MeV 
  seems to indicate that
the \pp data with polarized target are quite compatible with  the 
Crystal-Barrel and other LEAR data which need, in order to be explained,
a broad $f_0(1370)$ and a narrower $f_0(1500)$ 
\cite{pdg98}. In \cite{kll299} we have furthermore studied the
dependence of the positions of the $S$-matrix singularities on the
interchannel coupling strengths to find origin of resonances. We have also
looked at the interplay between $S$-matrix zeroes and poles.

\begin{table}[htb]
\caption{Average masses and widths of resonances}
\newcommand{\m}{\hphantom{$-$}}
\newcommand{\cc}[1]{\multicolumn{1}{c}{#1}}
\renewcommand{\tabcolsep}{2pc} 
\renewcommand{\arraystretch}{1.2} 
\begin{tabular}{@{}cccc}
\hline
resonance & mass (MeV) & width (MeV) & sheet \\
\hline 
\epsig or $\sigma$ & $523 \pm 12$ & $518 \pm 14$ & $-++$ \\
\hline
\fo & $991 \pm 3$ & $71 \pm 14$ &  $-++$ \\
\hline
& $1406 \pm 19$ & $160 \pm 12$ & $---$ \\
\epw & $1447  \pm 27$ & $108 \pm 46$ & $--+$ \\
\hline 
\end{tabular}
\label{resonances}
\end{table}

In the \pp channel ($j= 1$) one can define three
 branching ratios
$b_{1j} = \sigma_{1j}/\sigma^{tot}_{11}~, j=1,2,3$.
Our model has in total nine such ratios.
Here  $\sigma_{11}$ is the elastic \pp cross section, $\sigma_{1j}$
are the transition cross sections to 
\kk ($j=2$) and \roro ($j=3$)
and $\sigma^{tot}_{11}$ is the total \pp cross section.
One has $b_{11}+b_{12}+b_{13}=1.$
The energy dependence of these ratios is plotted in Fig.~1 for
solution B.


\begin{figure}[h]
 \vspace{6.cm}
\includegraphics{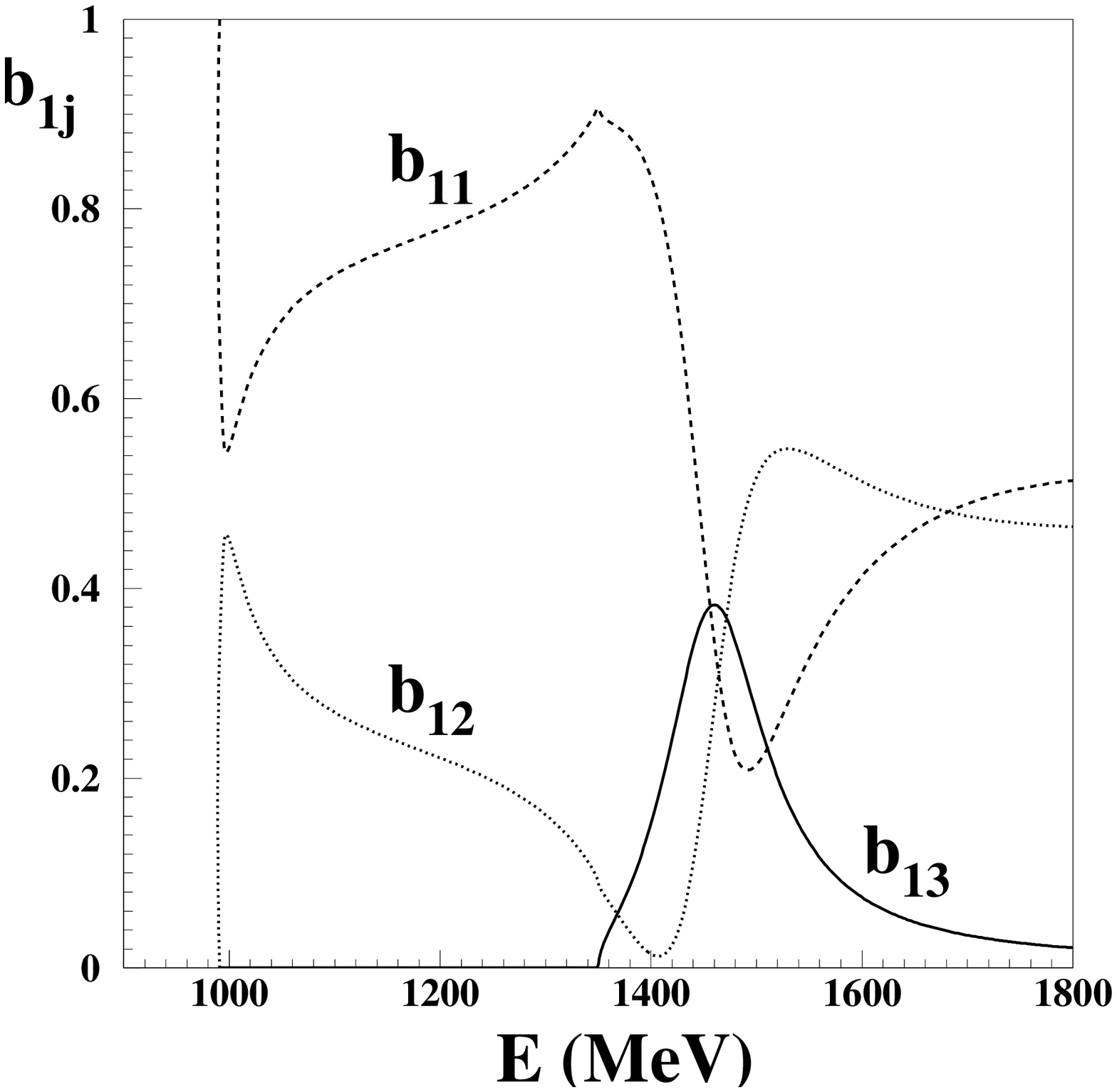}
\caption{
Branching ratios for \pp transitions to
 \pp $(b_{11})$, \kk $(b_{12})$ and \roro $(b_{13})$}
                \label{br}
  \end{figure}
Above the \kk threshold one can define an average branching ratio:
$$\overline b_{12}= \frac{1}{M_{max}-M_{min}}
\int_{M_{min}}^{M_{max}}b_{12}(E) dE.$$
In   \cite{amsler97} the branching ratios for the $f_0(1500)$ decay into
five channels, \pp, $\eta\eta$, $\eta\eta^{,}$, \kk and $4\pi$, are given as
29, 5, 1, 3 and 62 \%, respectively. The two main
disintegration channels are \pp and $4\pi$. In our model the $4\pi$ channel
is represented by the effective
\roro channel and we also obtain large fractions for the  averaged branching
 ratios $\overline{b}_{11}$ and $\overline{b}_{13}$. If we calculate the ratios
 $b_{13}/b_{11}$ exactly at 1500 MeV then we obtain numbers
 2.4, 1.2 and 2.3 for the solutions A, B and E, respectively. These numbers
 show the
 importance of the $4\pi$ channel in agreement with the experimental result of
 \cite{amsler97}. If we choose the energy interval from 
 1350 MeV to 1500 MeV our
 average branching ratios near $f_0(1400)$ for our solution B
 are $\overline{b}_{11}=0.61$ ,
 $\overline{b}_{12}=0.16$  and
 $\overline{b}_{13}=0.23$ .
We
 know that the extraction of the branching ratios from experiment is a
 difficult task \cite{abele96bis}.
 The average branching ratios depend quite
 sensitively on the energy bin chosen in the actual calculation as
 seen in Fig.~1. 
  In particular  the branching ratio
 $b_{12}$ (\pp$\rightarrow$\kk transition) is very
 small around 1420 MeV, close to the position of our \epw resonance poles.
 This is in qualitative agreement with the small number for the \kk
 branching ratio (3 \%) given in \cite{amsler97}.


\section{CROSSING SYMMETRY AND CHIRAL CONSTRAINTS}

We have looked for a new
solution fitting the previous \pp and \kk data and satisfying some chiral
constraints
at the \pp threshold  $s=4m_{\pi}^2$ ~\cite{donoghue90}. We have fixed 2
parameters of our model in such a way that the \pp scattering amplitude 
$T_{11}(4m_{\pi}^2) = 0.21m_{\pi}^{-1}$ and the \kk amplitude
$T_{22}(4m_{\pi}^2) = 0.13m_{\pi}^{-1}$.
The first value corresponds to a scalar-isoscalar scattering length $a_0^0$
close to those obtained in two loop calculations in chiral perturbation theory
\cite{bijnens97} and the second, for the \kk, to the leading order
value~\cite{donoghue90}.
Our unitary model for the $J=I=0$ amplitudes should be supplemented by a
suitable
parameterization of $J=0, I=2$ and $J=I=1$ waves in order to satisfy some
minimum crossing symmetry properties.
Parameters of our separable potentials can  be constrained in
such a way that the above set of amplitudes satisfies in an approximate way
Roy's equations~\cite{roy71}.
In order to do so we have used the equations
given in ~\cite{basdevant72} with the higher energy and $J \ge$ 2
contributions as
estimated in ~\cite{pennington73}. We have integrated the partial wave
spectral functions up to $s=46m_{\pi}^2$.
 The parameterization given in ~\cite{klr} has been used for the $I=J=1$ wave.
 For the scalar isotensor  wave we
have built a fit to the available phases~\cite{hoogland74}
 as in ~\cite{klr}
but with a rank 2 separable potential imposing a scattering length,
$a_0^2= -0.045m_{\pi}^{-1}$, close to the two loop results of
\cite{bijnens97}. With such a value the new set of the three
amplitudes satisfies better
 Roy's equations as can be seen in Fig.~2.
\begin{figure}[h]

\vspace{12.0 cm}
    
\includegraphics{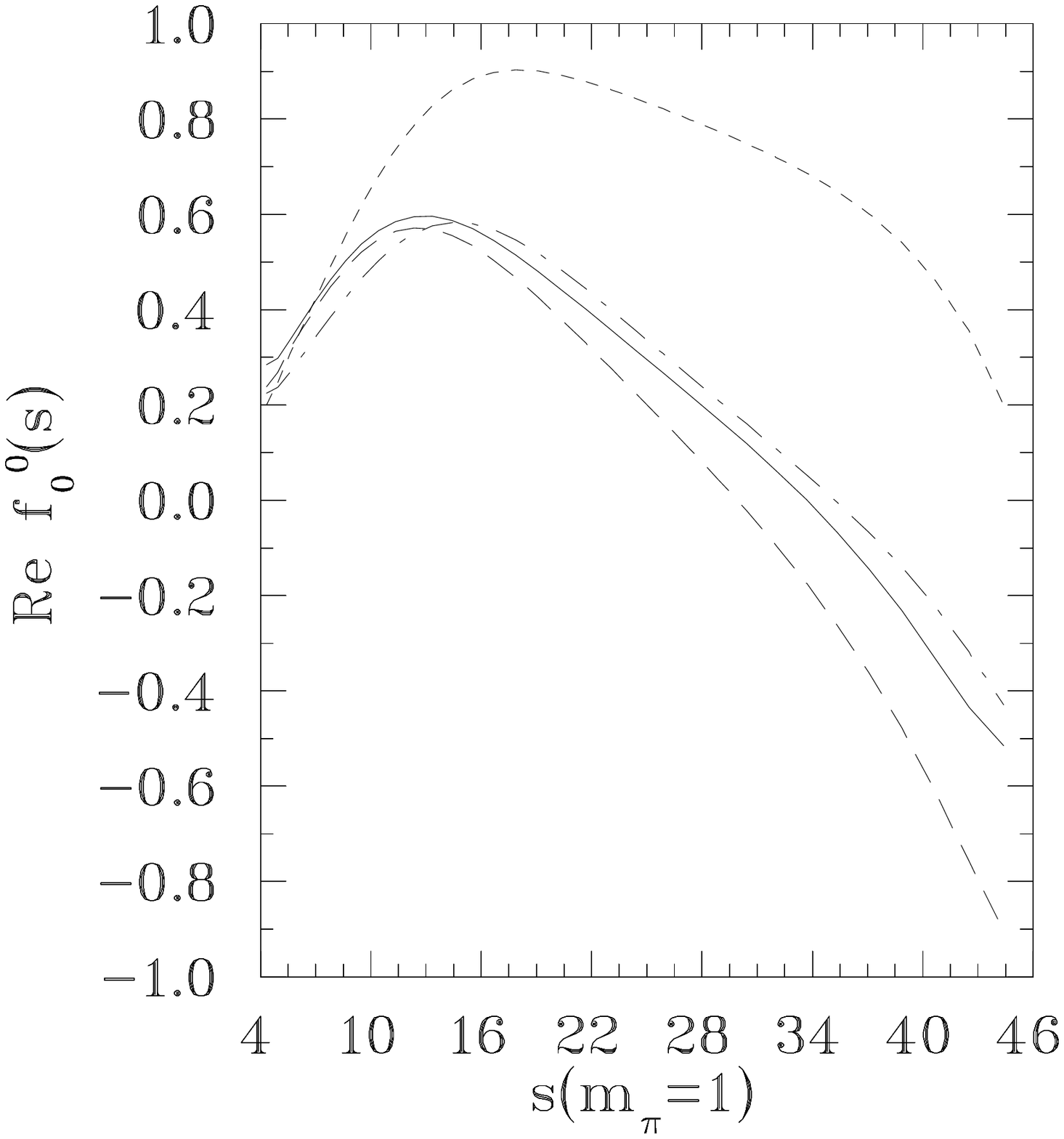}
\includegraphics{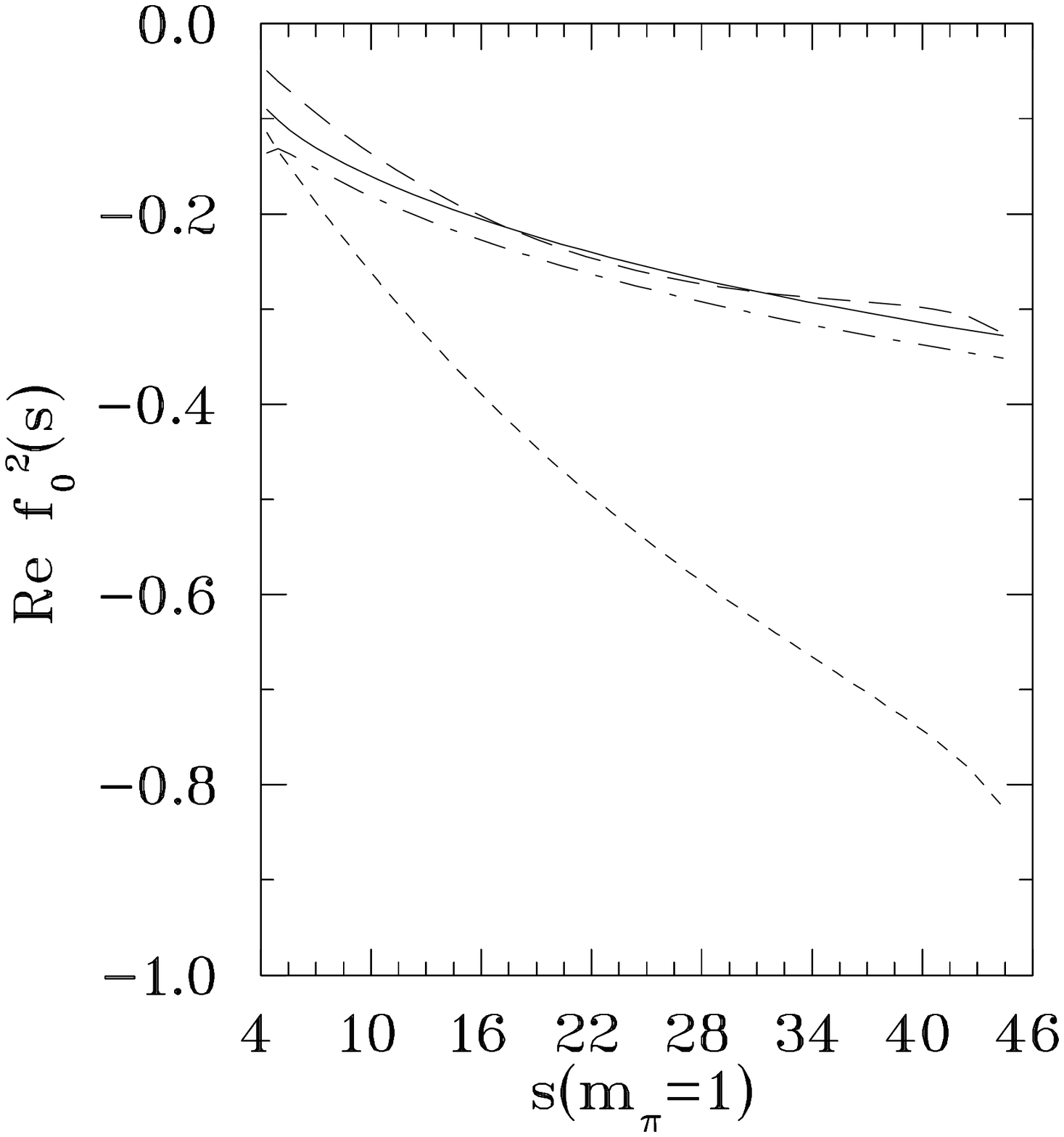}

\vspace{-0.4cm}

\caption{ Tests of Roy's equations for $Re f^0_0 (s)$ and $Re f^2_0(s)$
(see text)}
                \label{roy}
  \end{figure}
There we have
compared (for $J=0$ and $I=0,2$)  the real parts of the partial wave
amplitudes $Re f^I_J(s)$ as calculated from
$Re f^I_J = (1/2) \sqrt{s/(s-4)} sin 2 \delta^I_J(s)$
for solution A~\cite{kll297} (dash-dot line)  and for the
new solution (solid line) to those given by Roy's equations for the
solution A (short-dashed line) and for the new solution (long-dashed line).
We find that if  $a_0^0$ is close to 0.2 $m_{\pi}^{-1}$
then $a_0^2$ should be in the vicinity of -0.04 $m_{\pi}^{-1}$ in order 
to satisfy Roy's equations for
the isoscalar and isotensor waves.
The  $I=J=1$ wave, not shown here, is less sensitive to these values
and does satisfy Roy's equation relatively well.
This preliminary study 
can be further extended by inclusion of other possible
chiral constraints  such
as those on the transition \pp to \kk.
One can also try to improve treatment of high partial
waves and high energy contributions to Roy's equations.


%
%
\vspace{1mm}
{\it Acknowledgments:} We thank B. Moussallam, J. Stern and R. Vinh Mau 
for helpful
discussions. This work has been supported by IN2P3-Polish laboratories
Convention (project No. 99-97). R. Kami\'nski thanks NATO for a grant.

\end{document}